# Enhanced solid solution hardening by off-center substitutional solute atoms in α-Ti

Zi-Han Yu[1,2], Shuo Cao[1], Rui Yang[1], Qing-Miao Hu[1,*]

[1] *Institute of Metal Research, Chinese Academy of Sciences, Shenyang 110016, China*

[2] *School of Materials Science and Engineering, University of Science and Technology of China, Shenyang 110016, China*

**Abstract:**

Most recently, some substitutional solute atoms in α-Ti have been predicted to occupy unexpectedly the low-symmetry (LS) positions away from the high-symmetry (HS) lattice site, which was speculated to result in enhanced solid solution hardening (SSH). In the present work, the SSH induced by the LS off-center solute atom is evaluated within the framework of continuum elasticity theory, in comparison with that induced by its HS lattice-site counterpart. The interaction energy and force between the solute atom and the basal/prismatic edge/screw ⟨a⟩ dislocations in α-Ti solid solution are calculated with the elastic dipole model, with which the strength increments induced by the solute atoms are evaluated with the Labusch model. We show that, in general, the LS solute atom interacts much more strongly with the dislocations than its HS counterpart does. The calculated interaction energies suggest that the LS solute atom forms atmosphere above/below the slip plane of the basal ⟨a⟩ dislocations but on the slip plane of the prismatic ⟨a⟩ dislocations regardless of the dislocation types (edge or screw). The strength increments caused by most of the LS solute atoms are more than an order of magnitude higher than those by their HS counterparts. The SSH effect induced by the LS solute atom is mainly determined by the strength of the Jahn-Teller splitting of the *d*-orbitals of the solute atom, dissimilar to that induced by HS solute atom where the atomic size mismatch dominates.

---

[*] Corresponding author: qmhu@imr.ac.cn

## 1. Introduction

Solid solution hardening (SSH) [1] is the most commonly used approach to improve the strength of metals. SSH arises from various interactions between solute atoms (SA) and dislocations, among which elastic interaction plays the major role [2]. A SA introduced in a crystal distorts the local lattice around it such that local strain field or local lattice distortion (LLD) is resulted. When a dislocation approaches the SA, the stress field induced by the dislocation interacts with the local strain field around the SA. Such interaction may drag the motion of the dislocation, leading to the SSH effect.

According to the classical metal physics, a substitutional SA in a metal solid solution occupies the high-symmetry (HS) lattice site without changing the point group symmetry of the crystal lattice [3]. With which, the host atoms with the same distance to the SA shift identically away or toward the SA depending on the size mismatch between the SA and host atoms. Namely, the LLD induced by the SA is highly uniform. This allows the SA to be treated as an elastic sphere with radius $R_s$ accommodated in a spherical hole with radius $R_h$ in the metal matrix to calculate the interaction between the SA and the dislocation within the framework of elastic continuum media theory [3,4]. Namely, the SA is assumed to be an elastic monopole. Based on the above assumption, some classical models have been developed by Cottrell [5,6], Fleischer [7], and Labusch [8], et al., to evaluate the SSH effect in substitutional metal solid solutions. It is noted that, the elastic theory for the SA-dislocation interaction fails in the dislocation core region [3]. Nowadays, some advanced computational techniques such as dislocation dipole [9–14] and flexible boundary condition [15–19] have been adopted to address the interaction between the SA and the dislocation core. However, the SSH models within the framework of the elastic continuum media theory have been demonstrated to be reliable enough to reproduce the experimental critical resolved shear stress (CRSS) of the substitutional solid solutions such as Al-Cu/Mg/Ga/Zn [20], Ni-

Ta/W/Re [21], Cu-Si/Mg/In [22], Nb-Mo [23], Ti-Re [24], Mo-Hf [25].

Most recently, the commonsense that the substitutional SA occupies the HS lattice site was challenged for some of the solid solutions. Hu and Yang [26] found that, in Ti matrix with hexagonal close packed (hcp) structure, the transition metal (TM) SA in the middle of the Chemical Element Periodic Table (CEPT) shifts away from the HS lattice site to a low-symmetry (LS) off-center site such that the point group symmetry is broken spontaneously (see Fig. 1). The physical origin of the symmetry breaking is that the HS configuration is energetically unfavorable because the $d$ electronic orbitals of the SA at the HS lattice site are highly degenerated. When the SA shifts to the LS site, Jahn-Teller splitting of the degenerated $d$ orbitals occurs, which lowers the energy of the system and makes the LS configuration more stable. As a result of the LS occupation of the SA, the LLD in the LS configuration becomes highly non-uniform and much larger than that in the HS configuration [26,27]. In this case, the classical SSH model based on the uniform LLD assumption is not expected to work properly. Hu and Yang speculated that the LS-SA should contribute more to the SSH than that predicted by the classical SSH model [26], which needs to be confirmed.

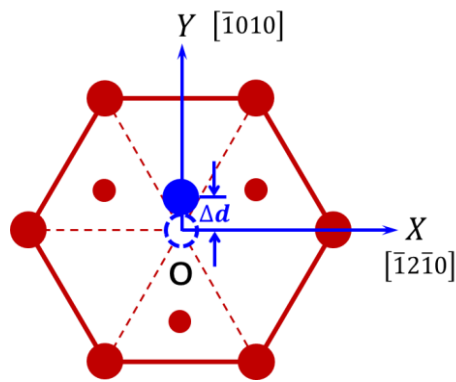

**Fig. 1.** Schematical representation of the off-center occupation of the solute atom in hexagonal close packed α-Ti. The atoms are projected to the (0001) basal plane of α-Ti. The large brown solid circles represent the host Ti atoms in the (0001) basal plane while the small ones represent the Ti atoms right above and below the plane. The open blue circle represents solute atom occupying the high-symmetry lattice site while the solid blue one is for the solute atom locating at the low-symmetry off-center site in the basal plane. $\Delta d$ is the in-plane distance between the high-symmetry lattice site and the low-symmetry off-center site.

According to Hu and Yang [26], the LS-SA is actually an orthorhombic point defect, which forms an elastic dipole. Therefore, the elastic dipole model [19,28–31] is required to evaluate the SSH effect induced by the LS-SA. The elastic dipole model has been used to calculate the interaction between the dislocation and the impurities such as C, N, O in the asymmetric octahedral interstitial site of α-Fe with body centered cubic (bcc) structure [32–35]. Limited by the textbook knowledge that the substitutional SA in metal solid solution is an elastic monopole, researchers have not yet realized that the SSH in some substitutional metal solid solution may need to be evaluated with the elastic dipole model as well.

The aim of the present work is to investigate how much the LS-SA contributes to the SSH of α-Ti solid solution as compared with the HS one within the framework of elastic continuum media theory. The local strains induced by the SAs on the HS and LS sites are determined through first principles calculations based on electronic density functional theory (DFT) [36]. Then, the interaction energy and force between the SAs and the dislocation are calculated by using the elastic dipole model. Subsequently, the SSH effects induced by the HS- and LS-SAs are evaluated with Labusch model [8]. Our work identifies some peculiar behavior of the formation of solute atmosphere around the dislocation. Meanwhile, we demonstrate that the contributions of the LS off-center SAs to the strength are much more significant than the HS lattice site ones, confirming our previous speculation. By the way, the interaction between the LS-SAs and the dislocation core calculated with the dislocation dipole model will be addressed in our following up work.

This paper is arranged as follows. In Section 2, we describe the method and calculation details of this work. In Section 3, we first examine the stable position (HS lattice site or LS off-center site) for the SAs in α-Ti. Then, the local strains induced by the SAs as well as the interaction energy and force between the solute atom and dislocation are evaluated and reported. The critical resolved shear stress (CRSS) increment due to the SSH is presented in Section 3 as well. The unique behavior of the LS off-center SA in α-Ti is discussed in Section 4. Finally, we conclude our work in Section 5.

## 2. Methodology and Calculation details

*2.1. Solid solution hardening model*

The SSH arises from the interaction between the stress field of the dislocation and the local strains or distortions induced by the SA. According to the isotropic elasticity theory, the stress fields of the edge and screw dislocations are respectively written as

$$\boldsymbol{\sigma}_E = \begin{pmatrix} \sigma_{xx} & \tau_{xy} & \tau_{xz} \\ \tau_{xy} & \sigma_{yy} & \tau_{yz} \\ \tau_{xz} & \tau_{yz} & \sigma_{zz} \end{pmatrix} = \frac{Gb}{2\pi(1-v)} \begin{pmatrix} \frac{-y(3x^2+y^2)}{(x^2+y^2)^2} & \frac{x(x^2-y^2)}{(x^2+y^2)^2} & 0 \\ \frac{x(x^2-y^2)}{(x^2+y^2)^2} & \frac{y(x^2-y^2)}{(x^2+y^2)^2} & 0 \\ 0 & 0 & \frac{-2vy}{x^2+y^2} \end{pmatrix} \quad (1)$$

and

$$\boldsymbol{\sigma}_S = \begin{pmatrix} \sigma_{xx} & \tau_{xy} & \tau_{xz} \\ \tau_{xy} & \sigma_{yy} & \tau_{yz} \\ \tau_{xz} & \tau_{yz} & \sigma_{zz} \end{pmatrix} = \frac{Gb}{2\pi} \begin{pmatrix} 0 & 0 & \frac{-y}{x^2+y^2} \\ 0 & 0 & \frac{x}{x^2+y^2} \\ \frac{-y}{x^2+y^2} & \frac{x}{x^2+y^2} & 0 \end{pmatrix} \quad (2)$$

with $G$ being the shear modulus, $b$ the Burgers vector, $v$ the Poisson's ratio. The stress field is expressed within the Cartesian coordinate system in terms of the slip system of the dislocation concerned, with $x$ being the slip direction of the dislocation, $y$ being the normal of the slip plane, and $z$ along the dislocation line (see Fig. 2). In the present work, we focus ourselves on the ⟨a⟩ dislocations with Burgers vector of $\frac{1}{3}\langle 11\bar{2}0 \rangle$ gliding on the (0001) basal plane (BPD, basal plane dislocation) and $(1\bar{1}00)$ prismatic plane (PPD, prismatic plane dislocation) because they are the main slip systems for the deformation of α-Ti alloys [3,37].

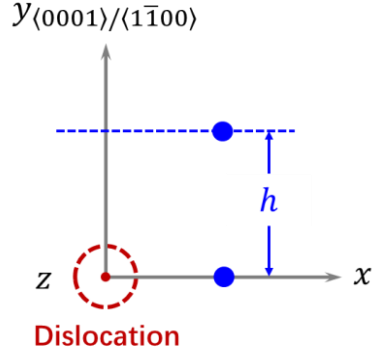

**Fig. 2.** Schematic representation of a dislocation and a solute atom in the Cartesian coordinate system defined with respect to the slip system. The $z$ axis of the Cartesian coordinate is along the dislocation line (vertical to the paper), which is $[1\bar{1}00]$ for the basal plane $\langle a \rangle$ edge dislocation, $[11\bar{2}0]$ for the basal plane $\langle a \rangle$ screw dislocation, $[0001]$ for the prismatic plane $\langle a \rangle$ edge dislocation, and $[11\bar{2}0]$ for the prismatic plane $\langle a \rangle$ screw dislocation. The $y$ axis is set as the normal of the slip plane, $y_{\langle 0001 \rangle}$ for basal plane slip and $y_{\langle 1\bar{1}00 \rangle}$ for prismatic plane slip. The dislocation moves along the $x$ axis. The blue dots are the solute atoms with the different height $h$ from the slip plane.

For substitutional metal solid solutions, the classical SSH model assumes that the SA is an elastic sphere with radius of $R_s$ accommodated in a hole of the matrix with radius of $R_h$. The radius mismatch (or local strain) is $\delta = (R_s - R_h)/R_h$ and the volume change $\Delta V = 4\pi \delta R_h^3$. The local strain induced by the SA is treated as an elastic monopole. The elastic interaction energy between the SA and the dislocation is thus expressed as

$$E^{\text{int}} = p \cdot \Delta V, \tag{3}$$

with the hydrostatic pressure

$$p = \frac{(1+v)Gb}{3\pi(1-v)} \cdot \frac{y}{x^2 + y^2}. \tag{4}$$

As introduced in Section 1, the SA at the LS off-center site is an orthorhombic defect such that the induced local strain should be treated as an elastic dipole. Therefore, the interaction between the LS-SA and dislocation may not be evaluated with the above elastic monopole SSH model. Instead, the elastic dipole model has to be adopted. The elastic dipole may be described as the second rank tensor ($\lambda$-tensor) [38]

$$\lambda_{ij} = \frac{\partial \varepsilon_{ij}}{\partial c_0}, \tag{5}$$

with $\varepsilon_{ij}$ being the component of the strain tensor $\boldsymbol{\varepsilon}$ caused by the SA and $c_0$ being the atomic fraction of the SA. $i,j = 1,2,3$ are the index for the matrix components. In principal axial coordinate system for the SA, the $\lambda$-tensor has only three non-zero components (the three principal components), i.e.,

$$\boldsymbol{\lambda} = \begin{pmatrix} \lambda_1 & 0 & 0 \\ 0 & \lambda_2 & 0 \\ 0 & 0 & \lambda_3 \end{pmatrix}. \tag{6}$$

Accordingly, for the α-Ti solid solution in Cartesian coordinate system, the three principal strain tensor components may be expressed as

$$\varepsilon_1 = \frac{X' - X}{X} \tag{7}$$

$$\varepsilon_2 = \frac{Y' - Y}{Y} \tag{8}$$

$$\varepsilon_3 = \frac{Z' - Z}{Z}. \tag{9}$$

Here, $(X, Y, Z)$ and $(X', Y', Z')$ are respectively the lattice constants of the pure host metal and the solid solution in Cartesian coordinates system (see Fig. 1), which may be obtained readily through first principles calculations.

With the calculated $\lambda$-tensor, the elastic interaction energy between the SA and the dislocation is expressed as [4,39]

$$E^{\text{int}} = -V_0 \lambda_{ij} \sigma_{ij}, \tag{10}$$

where $V_0$ is the atomic volume of the reference solid [4,20]. $\sigma_{ij} = \tau_{ij}$ for $i \neq j$. A positive value of $E^{\text{int}}$ represents repulsion between an SA and a dislocation and a negative one corresponds to attraction. It is noted that, the orientation of the principal axial coordinate system for the SA may be different from that of the coordinate system for the stress field of the dislocation. Therefore, to calculate the interaction energy with Eq. 10, we need to transform the $\lambda$-tensor in the principal axial coordinate system for the SA to $\lambda'$-tensor in the coordinate system for the stress field $\boldsymbol{\sigma}$ of the dislocations:

$$\lambda'_{ij} = a_{ik} a_{jl} \lambda_{kl} \quad (i,j = x, y, z; \ k, l = 1, 2, 3), \tag{11}$$

with $a_{ik}$ and $a_{jl}$ being the orientation cosine matrices between the coordinate system for the stress field and the principal axial coordinate system for the original $\lambda$-tensor.

With the obtained interaction energy between the SA and the dislocation $E^{\text{int}}$, the force acting on the dislocation by the SA may be calculated as [3]:

$$F_x = \frac{\partial E^{\text{int}}}{\partial x}. \tag{12}$$

with $x$ being the displacement of dislocation in Cartesian coordinate system along the glide direction of the dislocation. Because the dislocation core region is severely distorted and does not obey the elasticity theory, a cylinder with a hole of radius $r_0 = b \sim 4b$ is artificially set in the center of the dislocation model under the continuum medium approximation to avoid the failure of the elasticity theory [3].

According to Labusch model, the contribution of SSH to the critical resolved shear stress (CRSS) is evaluated as

$$\Delta \tau = \frac{1}{S_\text{F}} \left( \frac{F_m^4 c^2 w}{4Gb^9} \right)^{\frac{1}{3}}. \tag{13}$$

Here, $F_\text{m}$ is the maximum force acting on the dislocation by the SA. $c$ is the concentration of the SA. $w$ (~ $5b$ [40]) is the interaction range of SA and dislocation. $S_\text{F}$ is the Schmid factor.

*2.2. Calculations details*

To model the α-Ti solid solution, we construct supercell with size of $m \times m \times n$ ($m = 3, 4, 5$ and $n = 2, 3, 4$) times of the hcp unit cell, in which one Ti atom is replaced by an SA (SA = V, Nb, Ta, Cr, Mo, W, Mn, Tc, and Re). A first principles method is adopted to optimize fully the lattice parameters and atomic positions of the supercell through minimizing the stress and interatomic forces. We adopt two tactics to conduct the geometric optimization. One is that the geometric optimization starts from the supercell with SA placed at the HS lattice site (initial HS-SA configuration) such that the point group of the system maintains. In this case, SA should remain at the HS site after the geometric optimization because the total force acting on SA by the host atoms is exactly zero due to the point group symmetry of the system. The other tactic is that the SA in the initial supercell is manually shifted slightly away from the HS lattice site to break the point group of the system (initial LS-SA configuration). This allows the geometric optimization to find the possible stable LS off-center site for the SA. The SA favoring the HS lattice site should shift back to the HS lattice site after the geometric optimization. The lattice constants of the optimized supercells are taken to evaluate the $\lambda$-tensor (Eqs. 7~9).

The calculations are performed by using the first principles plane wave method implemented in the Vienna Ab initio Simulation Package (VASP) [41]. The project augmented wave potential [42] is employed for the interaction between the ionic core and valence electrons. The generalized gradient approximation (GGA) parameterized by Perdew, Burke, and Ernzerhof (PBE) [43] is adopted for the electronic exchange and correlation. The plane-wave cutoff energy is set as 500 eV. The Monkhorst-Pack *k*-point grid is used to sample the reduced Brillouin zone. The density of the *k* points is set as about 0.03Å$^{-1}$. The energy tolerance for the electronic minimization is set as $1\times10^{-6}$ eV/atom and the force tolerance for the geometric optimization is set as $1\times10^{-2}$ eV/Å.

## 3. Results

*3.1. Off-center occupation of solute atoms*

Our previous calculations demonstrated that the LS off-center site for the SA always locates in the basal plane of the hcp lattice along the $\langle 10\bar{1}0 \rangle$ direction [26,27]. The LS off-center occupation may be measured with the in-basal-plane deviation of the SA from the HS lattice site, $\Delta d$, as shown in Fig. 1. In Table 1, we list the deviation $\Delta d$ for the Ti-SA (SA = V, Nb, Ta, Cr, Mo, W, Mn, Tc, and Re) solid solutions, calculated with the initial LS-SA configuration. As seen in the Table 1, for SA = V, Nb, Ta, $\Delta d$ is essentially zero. Namely, these solute atoms, although initially shifted away from the HS lattice site to the LS off-center site, go back to the HS lattice site after the geometric optimization, i.e., they are stable to occupy the HS lattice site. For SA = Cr, Mo, W, Mn, Tc, and Re, we get sizable $\Delta d$ value, meaning that these solute atoms prefer the LS off-center site. For SA in the same group of the CEPT, $\Delta d$ decreases in general from the top to the bottom (i.e, from Cr/Mn to Mo/Tc to W/Re).

Also listed in Table 1 are the energy differences between the supercells with SA occupying HS and LS sites, $\Delta E = E^{\text{LS}} - E^{\text{HS}}$. Here, $E^{\text{HS}}$ is total energy of the supercell with the HS-SA configuration whereas $E^{\text{LS}}$ is the one with the LS-SA configuration. As seen in the table, $\Delta E$ is negligible for SA = V, Nb, and Ta. For SA = Cr, Mo, W, Mn, Tc, and Re, $\Delta E$ is negative, i.e., the LS configuration is lower in energy than the HS configuration. The calculated $\Delta E$ demonstrates that the SAs Cr, Mo, W, Mn, Tc, and Re prefer the LS off-center site in α-Ti whereas V, Nb, and Ta prefer the HS lattice site. As explained in our previous publications [26,27], the shift of the SA from the HS lattice site to the LS off-center site is ascribed to the Jahn-Teller splitting of the degenerated $d$ orbitals of the HS-SA. Discernably, $|\Delta E|$ measures the strength of the Jahn-Teller splitting. A larger $|\Delta E|$ corresponds to a stronger Jahn-Teller splitting. As seen in Table 1, the Jahn-Teller splitting for Mn, Tc, and Re is stronger than that for Cr, Mo, W. $|\Delta E|$ for V, Nb, Ta is nearly zero, i.e., Jahn-Teller splitting does not occur for these SAs.

**Table 1**

Deviation of solute atoms from the high-symmetry lattice sites $\Delta d$ and the energy difference between the high-symmetry lattice site and low-symmetry off-center configurations $\Delta E = E^{\text{LS}} - E^{\text{HS}}$ calculated with 4×4×2 supercell.

|  | Ti-V | Ti-Nb | Ti-Ta | Ti-Cr | Ti-Mo | Ti-W | Ti-Mn | Ti-Tc | Ti-Re |
| --- | --- | --- | --- | --- | --- | --- | --- | --- | --- |
| $\Delta d$ (Å) | 0.047 | 0.028 | 0.007 | 0.344 | 0.319 | 0.248 | 0.439 | 0.373 | 0.322 |
| $\Delta E$ (eV) | 0.001 | 0.002 | -0.000 | -0.065 | -0.109 | -0.045 | -0.226 | -0.311 | -0.296 |

*3.2. Lattice constants and λ-tensor*

The SA induced **λ**-tensor is critical in the evaluation of the interaction between the SA and dislocation within the framework of elastic dipole model. According to Eq. 5, the components of **λ**-tensor may be determined by fitting the $\varepsilon_{ij} \sim c_0$ relationship, which requires to calculate the variation of the lattice constants against the SA concentration $c_0$. In the present work, the variation of $c_0$ is implemented by using supercells with various sizes containing only one substitutional SA.

Here, we take the Ti-Mo system as an example to show $\varepsilon_{ij} \sim c_0$ relationship. In Table 2, we list the lattice constants of pure Ti and Ti-Mo calculated with different supercell sizes. For the pure Ti and Ti-Mo supercells with the HS-Mo configuration, $Y'/X' = 0.866 \approx \sqrt{3}/2$ for all the supercell size, i.e., the crystal lattice remains hexagonal. For Ti-Mo supercell with the LS-Mo configuration, $Y'/X' < 0.866$, indicating that the system deviates from hexagonal and the LS-Mo forms an orthorhombic defect.

**Table 2**

Lattice constants (unit in Å) of the supercells of pure Ti and Ti-Mo with different sizes, in terms of the Cartesian coordinates system.

| Supercell | | 3×3×2 | 4×4×2 | 4×4×3 | 5×5×3 | 5×5×4 |
|---|---|---|---|---|---|---|
| $c_0$ | | 1/36 | 1/64 | 1/96 | 1/150 | 1/200 |
| Pure Ti | X | 8.82 | 11.75 | 11.75 | 14.68 | 14.69 |
|  | Y | 7.64 | 10.17 | 10.17 | 12.71 | 12.72 |
|  | Z | 9.29 | 9.30 | 13.95 | 13.95 | 18.60 |
| HS-Mo | X′ | 8.79 | 11.74 | 11.74 | 14.67 | 14.68 |
|  | Y′ | 7.62 | 10.17 | 10.17 | 12.71 | 12.72 |
|  | Z′ | 9.30 | 9.30 | 13.94 | 13.95 | 18.60 |
| LS-Mo | X′ | 8.95 | 11.85 | 11.81 | 14.72 | 14.71 |
|  | Y′ | 7.52 | 10.10 | 10.13 | 12.69 | 12.70 |
|  | Z′ | 9.25 | 9.27 | 13.92 | 13.93 | 18.58 |

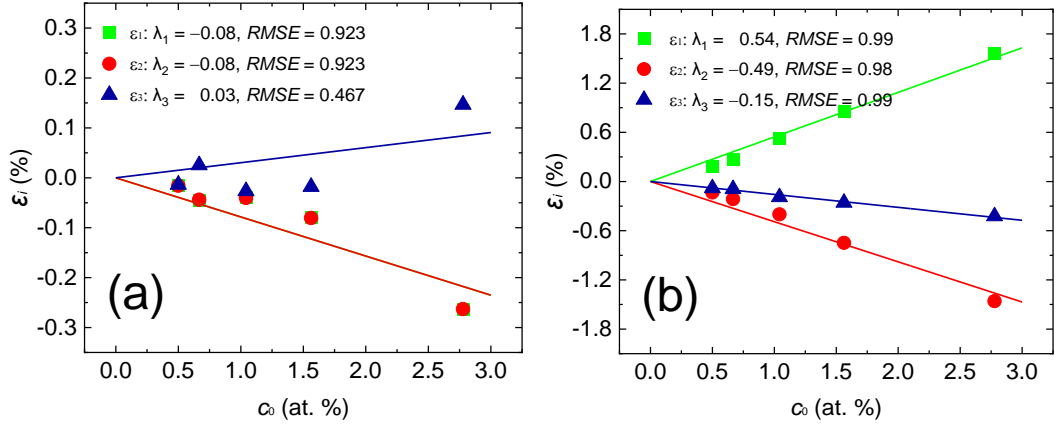

**Fig. 3.** Principal components of the strain tensor $\varepsilon_i$ as functions of the concentration of Mo atom at high-symmetry lattice site (a) and low-symmetry off-center site (b). The lines are the linear regressions of the scattered data points.

The strain tensor components $\varepsilon_i$ are evaluated for each of the supercell size and the $\varepsilon_i \sim c_0$ data points are plotted in Fig. 3. For both the HS-Mo (Fig. 3a) and LS-Mo (Fig. 3b) configurations, the absolute values of $\varepsilon_i$ increases in general with $c_0$. It is seen that the $\varepsilon_i \sim c_0$ relationship for the LS-Mo configuration is almost perfectly linear. However, it deviates from linear significantly for the HS-Mo configuration. The reason is that, for the HS-Mo configuration, the absolute value of $\varepsilon_i$ is very small. For the highest $c_0$ ($1/36 \approx 2.78\%$), the absolute value of $\varepsilon_i$ is no more than 0.3%. Therefore, the computational uncertainties in $\varepsilon_i$ is significant as compared to its absolute value, which matters in the fitting. For the LS-Mo configuration, the absolute value of $\varepsilon_i$ is large such that the uncertainties do not influence the fitting significantly. From the linear fit of the $\varepsilon_i \sim c_0$ relationship, we obtain the principal $\boldsymbol{\lambda}$-tensor components of $\lambda_1 = \lambda_2 = -0.08$ and $\lambda_3 = 0.03$ for the HS-Mo configuration and $\lambda_1 = 0.54$, $\lambda_2 = -0.49$, and $\lambda_3 = -0.15$ for the LS-Mo configuration.

Because $\varepsilon_i$ is proportional to $c_0$ as seen in Fig. 3b, $\lambda_i$ may be approximately taken as $\lambda_i = \varepsilon_i/c_0$. This approximation avoids the heavy load needed to calculate the strain tensors with various large supercells. Therefore, we adopt only 4×4×2 supercell to calculate $\lambda_i$ for all the other SAs. In Fig. 4, we display the principal components of the $\boldsymbol{\lambda}$-tensor of α-Ti-SA (SA = V, Nb, Ta, Cr, Mo, W, Mn, Tc, and Re) solid solution. A negative/positive value of $\lambda_i$ means that the lattice shrinks/expands along the $i$ axis.

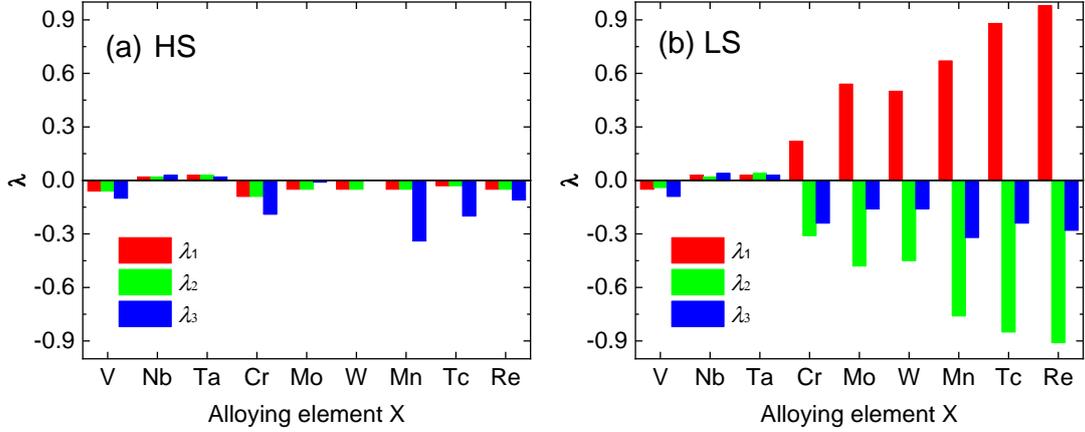

**Fig. 4.** Principal components of $\lambda$-tensor of α-Ti-SA (SA = V, Nb, Ta, Cr, Mo, W, Mn, Tc, and Re) solid solution calculated with the 4×4×2 supercell. The $\lambda$-tensors for the HS (a) and LS configurations (b) are evaluated respectively using supercells optimized with initial HS-SA configuration (i.e., starting with SA placed on the high-symmetry lattice site) and initial LS-SA configuration (i.e., starting with SA manually shifted away from the high-symmetry lattice site).

As seen in Fig. 4a, for the HS-SA configuration, $\lambda_1 = \lambda_2$ with an absolute value smaller than 0.10 for all the SAs, indicating that these alloying elements shrinks (for V, Cr, W, Mn Tc, and Re) or expand (for Nb and Ta) the lattice equally and slightly along both the $X$ and $Y$ axes. The lattice also shrinks or expands slightly along the $Z$ axis by SA = Nb, Ta, Mo, and W as shown by the corresponding small absolute value of $\lambda_3$. However, $\lambda_3$ is significantly negative for SA = V, Cr, Mn, Tc, and Re, meaning that the lattice shrinks significantly along the $Z$ axis.

As shown in Fig. 4b, for the LS-SA configuration, $\lambda_1$ and $\lambda_2$ remain very small and almost equal for SA = V, Nb, and Ta, because these solute atoms initially shifted away from the HS site go back to the HS site after the geometric optimization. However, for SA = Cr, Mo, W, Mn, Tc, and Re, we get sizable positive $\lambda_1$ and negative $\lambda_2$, indicating the lattice significantly expands along the $X$ axis and shrinks along the $Y$ axis. For SA = Nb and Ta, we obtain slightly positive $\lambda_3$. $\lambda_3$ for the other SAs is negative with quite large an absolute value.

The $\lambda$-tensor in the principal axis coordinate system (Fig. 1) of the elastic dipole is then transformed to the $\lambda'$-tensor in the Cartesian coordinate system defined with respect to the slip system (Fig. 2) according to Eq. 11. In Table 3, we list the $\lambda'$-tensor components for SAs occupying both the HS and LS sites for different dislocation slip systems including basal plane edge dislocation (BPED), basal plane screw dislocation (BPSD), prismatic plane edge dislocation (PPED), and prismatic plane screw dislocation (PPSD).

**Table 3**

The $\lambda'$-tensor components for solute atoms SA (SA = V, Nb, Ta, Cr, Mo, W, Mn, Tc, and Re) occupying the HS lattice site and the LS off-center site with respect to various dislocation types. If not listed, the value for the corresponding $\lambda'$-tensor component is zero.

| | SA | V | Nb | Ta | Cr | Mo | W | Mn | Tc | Re |
|---|---|---|---|---|---|---|---|---|---|---|
| \multicolumn{11}{l}{Basal plane edge dislocation (BPED) $[11\bar{2}0](0001)$} |
| HS | $\lambda'_{xx}$ | -0.036 | 0.012 | 0.018 | -0.054 | -0.030 | -0.030 | -0.030 | -0.018 | -0.030 |
| | $\lambda'_{yy}$ | -0.100 | 0.030 | 0.020 | -0.190 | -0.010 | 0.000 | -0.340 | -0.200 | -0.110 |
| | $\lambda'_{zz}$ | -0.084 | 0.028 | 0.042 | -0.126 | -0.070 | -0.070 | -0.070 | -0.042 | -0.070 |
| | $\lambda'_{xz}$ | 0.012 | -0.004 | -0.006 | 0.018 | 0.010 | 0.010 | 0.010 | 0.006 | 0.010 |
| LS | $\lambda'_{xx}$ | -0.029 | 0.017 | 0.019 | 0.079 | 0.222 | 0.205 | 0.259 | 0.355 | 0.399 |
| | $\lambda'_{yy}$ | -0.090 | 0.040 | 0.030 | -0.240 | -0.160 | -0.150 | -0.320 | -0.240 | -0.280 |
| | $\lambda'_{zz}$ | -0.061 | 0.033 | 0.051 | -0.169 | -0.162 | -0.155 | -0.349 | -0.325 | -0.329 |
| | $\lambda'_{xz}$ | 0.013 | -0.009 | -0.003 | -0.203 | -0.414 | -0.385 | -0.563 | -0.695 | -0.763 |
| \multicolumn{11}{l}{Prismatic plane edge dislocation (PPED) $[11\bar{2}0](1\bar{1}00)$} |
| HS | $\lambda'_{xx}$ | -0.036 | 0.012 | 0.018 | -0.054 | -0.030 | -0.030 | -0.030 | -0.018 | -0.030 |
| | $\lambda'_{yy}$ | -0.084 | 0.028 | 0.042 | -0.126 | -0.070 | -0.070 | -0.070 | -0.042 | -0.070 |
| | $\lambda'_{zz}$ | -0.100 | 0.030 | 0.020 | -0.190 | -0.010 | 0.000 | -0.340 | -0.200 | -0.110 |
| | $\lambda'_{xy}$ | 0.012 | -0.004 | -0.006 | 0.018 | 0.010 | 0.010 | 0.010 | 0.006 | 0.010 |
| LS | $\lambda'_{xx}$ | -0.029 | 0.017 | 0.019 | 0.079 | 0.222 | 0.205 | 0.259 | 0.355 | 0.399 |
| | $\lambda'_{yy}$ | -0.061 | 0.033 | 0.051 | -0.169 | -0.162 | -0.155 | -0.349 | -0.325 | -0.329 |
| | $\lambda'_{zz}$ | -0.090 | 0.040 | 0.030 | -0.240 | -0.160 | -0.150 | -0.320 | -0.240 | -0.280 |
| | $\lambda'_{xy}$ | 0.013 | -0.009 | -0.003 | -0.203 | -0.414 | -0.385 | -0.563 | -0.695 | -0.763 |
| \multicolumn{11}{l}{Basal plane screw dislocation (BPSD) $[11\bar{2}0](0001)$} |
| HS | $\lambda'_{xx}$ | -0.084 | 0.028 | 0.042 | -0.126 | -0.070 | -0.070 | -0.070 | -0.042 | -0.070 |
| | $\lambda'_{yy}$ | -0.100 | 0.030 | 0.020 | -0.190 | -0.010 | 0.000 | -0.340 | -0.200 | -0.110 |
|    | $\lambda'_{zz}$ | -0.036 | 0.012  | 0.018  | -0.054 | -0.030 | -0.030 | -0.030 | -0.018 | -0.030 |
|    | $\lambda'_{xz}$ | 0.012  | -0.004 | -0.006 | 0.018  | 0.010  | 0.010  | 0.010  | 0.006  | 0.010  |
|    | $\lambda'_{xx}$ | -0.061 | 0.033  | 0.051  | -0.169 | -0.162 | -0.155 | -0.349 | -0.325 | -0.329 |
| LS | $\lambda'_{yy}$ | -0.090 | 0.040  | 0.030  | -0.240 | -0.160 | -0.150 | -0.320 | -0.240 | -0.280 |
|    | $\lambda'_{zz}$ | -0.029 | 0.017  | 0.019  | 0.079  | 0.222  | 0.205  | 0.259  | 0.355  | 0.399  |
|    | $\lambda'_{xz}$ | 0.0130 | -0.009 | -0.003 | -0.203 | -0.414 | -0.385 | -0.563 | -0.695 | -0.763 |
| Prismatic plane screw dislocation (PPSD) $[11\bar{2}0](1\bar{1}00)$ | | | | | | | | | | |
|    | $\lambda'_{xx}$ | -0.100 | 0.030  | 0.020  | -0.190 | -0.010 | 0.000  | -0.340 | -0.200 | -0.110 |
| HS | $\lambda'_{yy}$ | -0.084 | 0.028  | 0.042  | -0.126 | -0.070 | -0.070 | -0.070 | -0.042 | -0.070 |
|    | $\lambda'_{zz}$ | -0.036 | 0.012  | 0.018  | -0.054 | -0.030 | -0.030 | -0.030 | -0.018 | -0.030 |
|    | $\lambda'_{yz}$ | 0.012  | -0.004 | -0.006 | 0.018  | 0.010  | 0.010  | 0.010  | 0.006  | 0.010  |
|    | $\lambda'_{xx}$ | -0.090 | 0.040  | 0.030  | -0.240 | -0.160 | -0.150 | -0.320 | -0.240 | -0.280 |
| LS | $\lambda'_{yy}$ | -0.061 | 0.033  | 0.051  | -0.169 | -0.162 | -0.155 | -0.349 | -0.325 | -0.329 |
|    | $\lambda'_{zz}$ | -0.029 | 0.017  | 0.019  | 0.079  | 0.222  | 0.205  | 0.259  | 0.355  | 0.399  |
|    | $\lambda'_{yz}$ | 0.013  | -0.009 | -0.003 | -0.203 | -0.414 | -0.385 | -0.563 | -0.695 | -0.763 |

### 3.3. Stress field of the dislocation

Within the framework of the elasticity theory, the stress field of the dislocation is determined by the shear modulus $G$, Burgers vector $b$, and Poisson's ratio $v$ (Eqs. 1 and 2). The calculated $G$, $b$, and $v$ are respectively 51.9 GPa, 2.94 Å, and 0.30, in good agreement with those from experimental measurements and other theoretical calculations [44,45]. Fig. 5 displays the stress field components of the dislocation. It is noted that $G$, $b$, and $v$ for BPD are the same as those for PPD. Therefore, the stress fields of BPD and PPD are identical. Here, $\sigma_{ij}$ with $i = j$ ($i,j = x,y,z$) is the normal stress component while $\sigma_{ij} = \tau_{ij}$ with $i \neq j$ is the shear stress component. It is seen that, for the edge dislocation, $\sigma_{xx}$ and $\sigma_{zz}$ are dipole shaped along the $y$ direction. The dipole of $\sigma_{xx}$ is irregular. $\sigma_{yy}$ and $\tau_{xy}$ are hexapole shaped with a strong dipole and a weak quadrupole. The strong dipoles for $\sigma_{yy}$ and $\tau_{xy}$ line respectively along the $y$ and $x$ directions. For the screw dislocation, both $\tau_{xz}$ and $\tau_{yz}$ are dipole shaped but with different orientations. According to Eq. 10, the $\boldsymbol{\lambda'}$-tensor components determine the contributions of the corresponding stress tensor components to the interaction energy between the SA and dislocation. The $\boldsymbol{\lambda'}$-tensor components (Table 3) may vary for the HS and LS site occupations and different SAs, resulting in various patterns of the interaction energy, which will be presented in the next subsection.

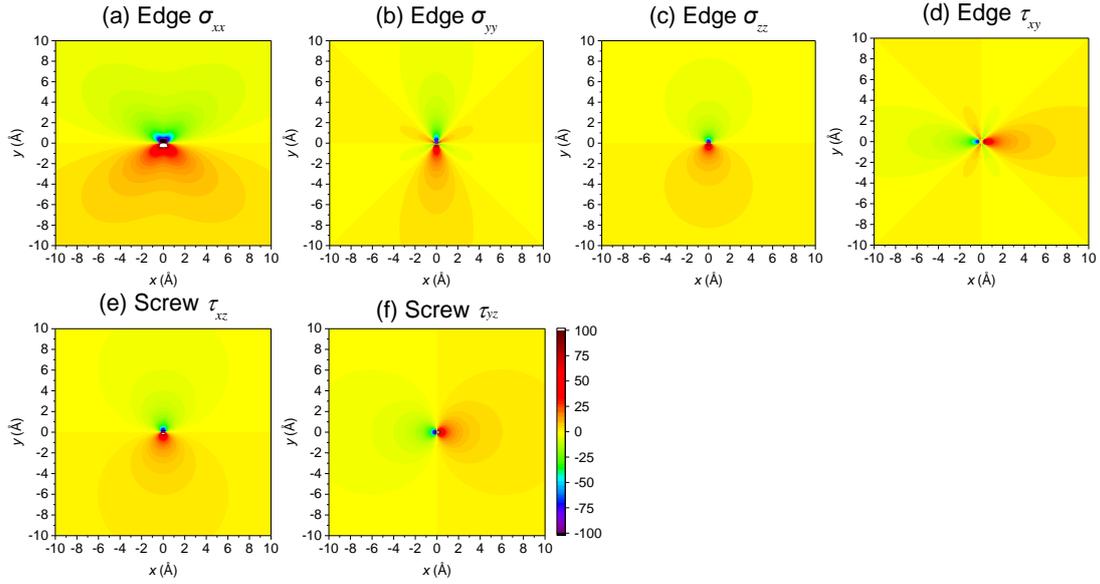

**Fig. 5.** Stress field of the dislocations. Subfigures (a), (b), (c), and (d) are respectively for the normal stresses $\sigma_{xx}$, $\sigma_{yy}$, $\sigma_{zz}$ and shear stress $\tau_{xy}$ of the edge dislocation while subfigures (e) and (f) are respectively for the shear stresses $\tau_{xz}$ and $\tau_{yz}$ of the screw dislocation.

*3.4. Interaction energy between the solute atom and dislocation*

With the calculated $\boldsymbol{\lambda}'$-tensor induced by the SA in α-Ti (subsection 3.2) and the stress tensor of the dislocations (subsection 3.3), the interaction energies $E^{\text{int}}(x, y)$ between SA and the dislocations are evaluated with Eqs. 10 and 11, which are presented as follows.

3.4.1. Interactions between SAs and edge dislocations

Fig. 6 displays the interaction energies between BPED and SA at HS lattice site (from Fig. 6a to Fig. 6i) and LS off-center site (from Fig. 6a′ to 6i′). The SA-dislocation interaction energy $E^{\text{int}}(x, y)$ for the HS-V/Nb/Ta is essentially the same as that for the LS-V/Nb/Ta because the initial LS-V/Nb/Ta configuration returns to the HS-VA/Nb/Ta configuration after the geometric optimization, leading to similar LS and HS $\boldsymbol{\lambda}'$-tensor components. The interaction between the LS/HS-V/Nb/Ta and BPED is very weak. In general, the interaction between the LS-Cr/Mo/W/Mn/Tc/Re and the BPED is slightly stronger than that between HS-Cr/Mo/W/Mn/Tc/Re and the BPED, except for Cr and Mn.

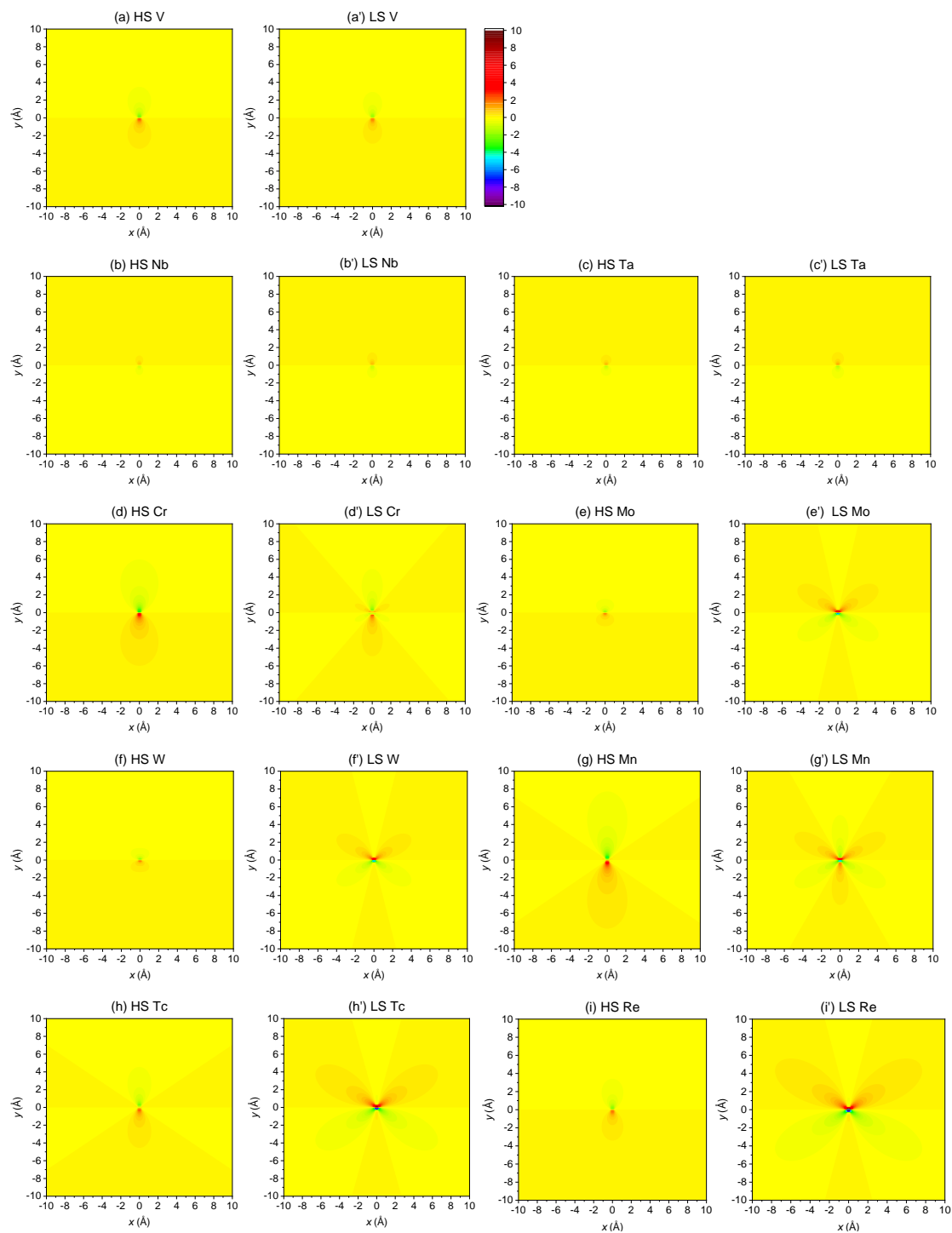

**Fig. 6.** Interaction energy (unit in eV) $E^{\text{int}}(x, y)$ between the solute atoms and basal plane ⟨a⟩ edge dislocations in α-Ti alloys. The subfigures from (a) to (i) are for SA = V, Nb, Ta, Cr, Mo, W, Mn, Tc, Re located at the high symmetry lattice site while those from (a′) to (i′) are for the SAs at the low symmetry off-center site.

For all the HS-SAs, the strongest SA-dislocation interaction, either attraction with negative $E^{\text{int}}(x,y)$ or repulsion with positive $E^{\text{int}}(x,y)$, occurs right below and above the dislocation line at $x = 0$. The HS-V/Cr/Mo/W/Mn/Tc/Re tend to accumulate above the slip plane because over there they are attracted by the dislocation. In contrast, the HS-Nb/Ta segregate below the slip plane as the attraction occurs below it. The accumulation of the SA above or below the slip plane of the edge dislocation is the well-known Cottrell atmosphere [5,6]. The reason that the HS-V/Cr/Mo/W/Mn/Tc/Re segregate above the slip plane whereas HS-Nb/Ta segregate below it is that the HS-V/Cr/Mo/W/Mn/Tc/Re shrink the host lattice whereas HS-Nb/Ta expand it as shown by the strains they induced (Fig. 4a).

The LS-Cr still accumulates right above the dislocation line as the HS-Cr does. However, the LS-Mo/W/Mn/Tc/Re exhibit segregation behavior very different from their HS counterparts. The LS-Mo/W/Tc/Re tend to accumulate bi-diagonally but not right below the dislocation line. Interestingly, LS-Mn shows clover type of segregation around the dislocation. It accumulates almost equivalently above and bi-diagonally below the dislocation line.

Fig. 7 shows the interaction energy $E^{\text{int}}(x,y)$ between the SA and the PPED. Similar to the case for the BPED, the interaction energies for HS-V/Nb/Ta and LS-V/Nb/Ta are very small and are almost the same because the optimized LS configuration is identical to the HS one. The strengths of $E^{\text{int}}(x,y)$ for LS-Cr and HS-Cr are close to each other. However, $E^{\text{int}}(x,y)$s for LS-Mo/W/Mn/Tc/Re are much higher than those for HS-Mo/W/Mn/Tc/Re.

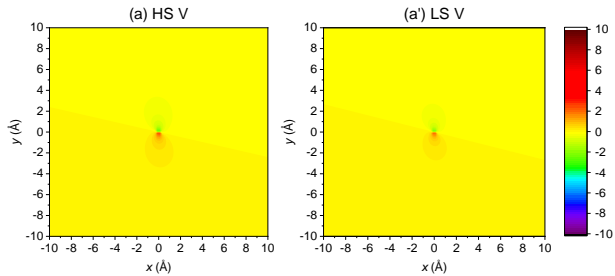

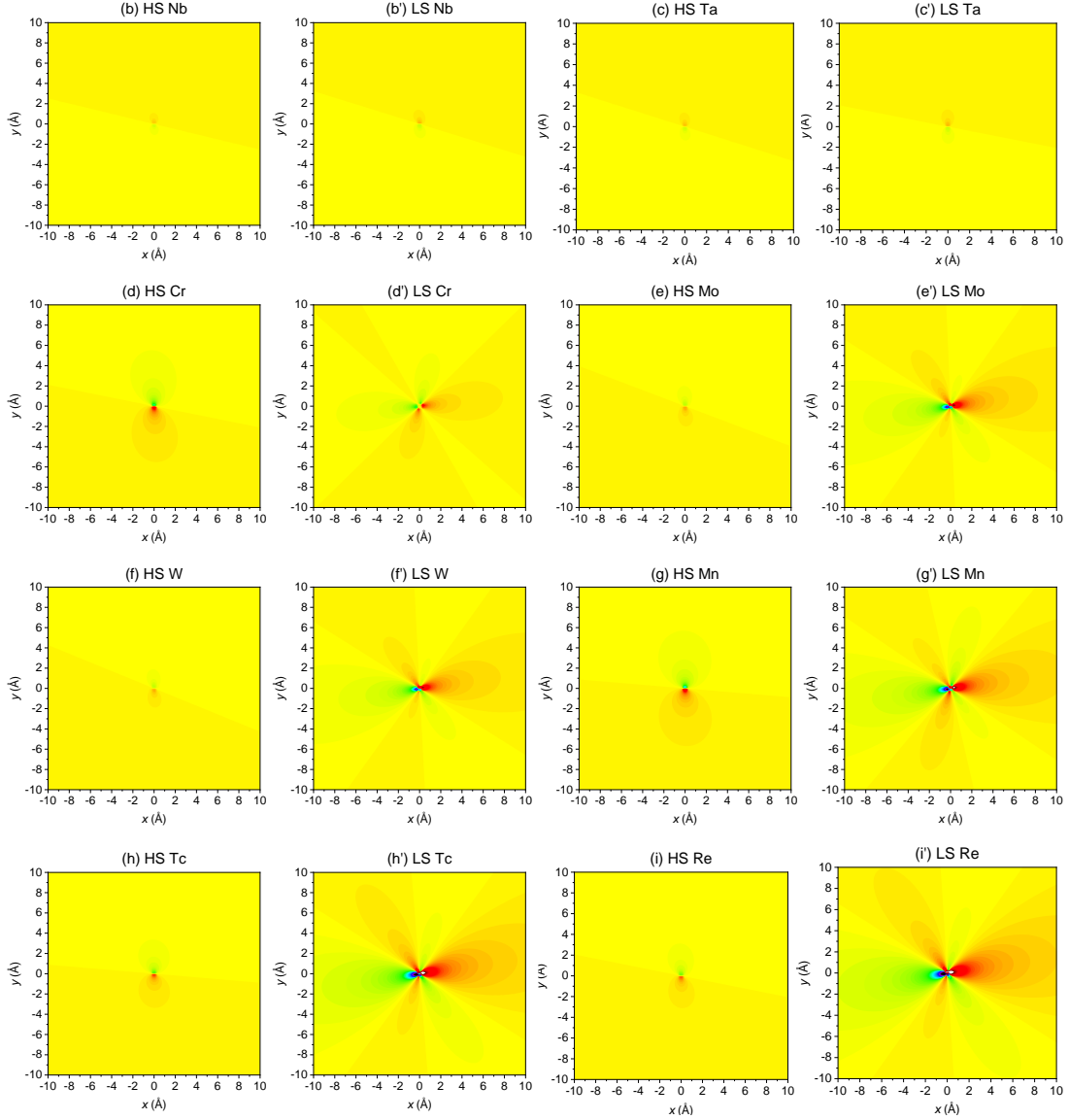

**Fig. 7.** Interaction energy (unit in eV) $E^{\text{int}}(x,y)$ between the solute atoms and prismatic plane ⟨a⟩ edge dislocations in α-Ti alloys. The subfigures from (a) to (i) are for SA = V, Nb, Ta, Cr, Mo, W, Mn, Tc, Re located at the HS lattice site while those from (a′) to (i′) are for the SAs at the LS off-center site.

Again, all the HS-SAs are attracted by the dislocation either right above (for V, Cr, Mo, W, Mn, Tc, and Re) or below (for Nb and Ta) the dislocation. Namely, the HS-SAs tend to segregate to the place right above or below the dislocation as usual, forming the Cottrell atmosphere. However, for the LS-Cr/Mo/W/Mn/Tc/Re, the strongest attraction appears nearly on the slip plane at the $x<0$ side, meaning that the LS-Cr/Mo/W/Mn/Tc/Re are apt to segregate almost on the slip plane of PPED.

3.4.2. Interactions between SAs and screw dislocations

Now, let's turn to the interaction between SA and screw ⟨a⟩ dislocation, which may be evaluated with Eqs. 10 and 11. As seen in Table 3, the $\boldsymbol{\lambda'}$-tensor components $\lambda'_{xz}$ and $\lambda'_{yz}$ of the HS-SAs are close to zero for both the BPSD and PPSD such that the shear stress components $\tau_{xz}$ and $\tau_{yz}$ contribute negligibly to $E^{\text{int}}(x,y)$. The normal stress components are exactly zero for the screw dislocation. Therefore, the interaction energy between the HS-SAs and the screw dislocations are negligible. Thus, we will not discuss them hereafter.

Fig. 8 displays the interaction energy $E^{\text{int}}(x,y)$ between the LS-SA and BPSD. It is seen that $E^{\text{int}}(x,y)$ is negligible for LS-V/Nb/Ta (same as their HS-SAs) because they go back to the HS site after the first principles geometric optimization. For LS-Cr/Mo/W/Mn/Tc/Re, $E^{\text{int}}(x,y)$ is dipole shaped along the $y$ direction. The interaction becomes stronger in the sequence of Cr, W, Mo, Mn, Tc, Re, in accordance with the absolute value of $\lambda'_{xz}$ of these SAs. The calculated interaction energy indicates that all the LS-SAs tend to segregate above the slip plane of the BPSD as shown by the negative interaction energy for $y > 0$.

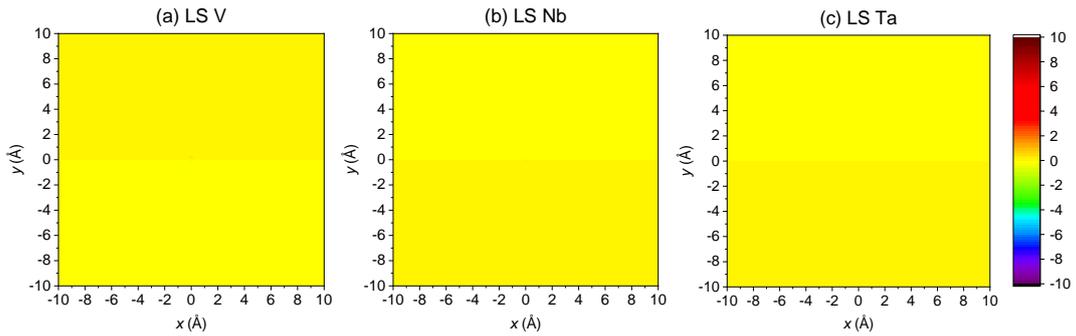

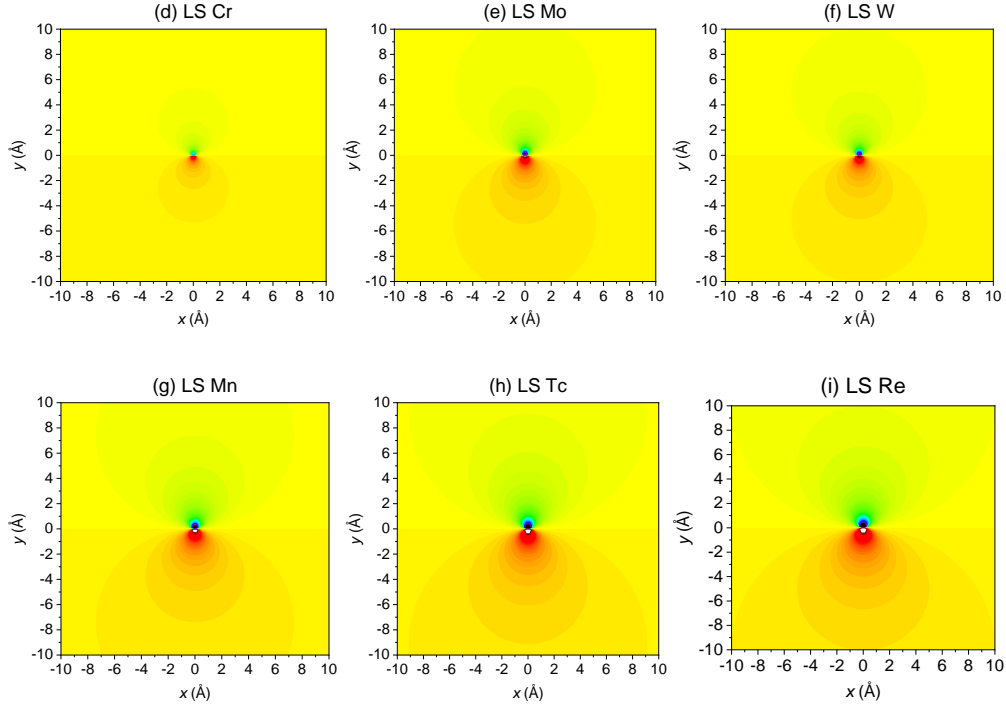

**Fig. 8.** Interaction energy (unit in eV) $E^{\text{int}}(x, y)$ between the solute atoms and basal plane ⟨a⟩ screw dislocations in α-Ti. The subfigures (a) to (i) are respectively for V, Nb, Ta, Cr, Mo, W, Mn, Tc, Re. The solute atoms are located at the LS off-center sites.

Fig. 9 displays the interaction energy $E^{\text{int}}(x, y)$ between the LS-SA and PPSD. Again, $E^{\text{int}}(x, y)$ is negligible for LS-V/Nb/Ta as they are actually HS-SAs. For LS-Cr/Mo/W/Mn/Tc/Re, $E^{\text{int}}(x, y)$ is dipole shaped along the $x$ direction. The interaction becomes stronger in the sequence of Cr, W, Mo, Mn, Tc, Re, in accordance with the absolute value of $\lambda'_{yz}$ of these SAs. Actually, the interaction energy between the LS-SA and PPSD is the same as that between the LS-SA and BPSD except that the directions of the dipole are different. According to the calculated interaction energy, the LS-SAs tend to segregate on the slip plane of the PPSD, ahead or behind the dislocation.

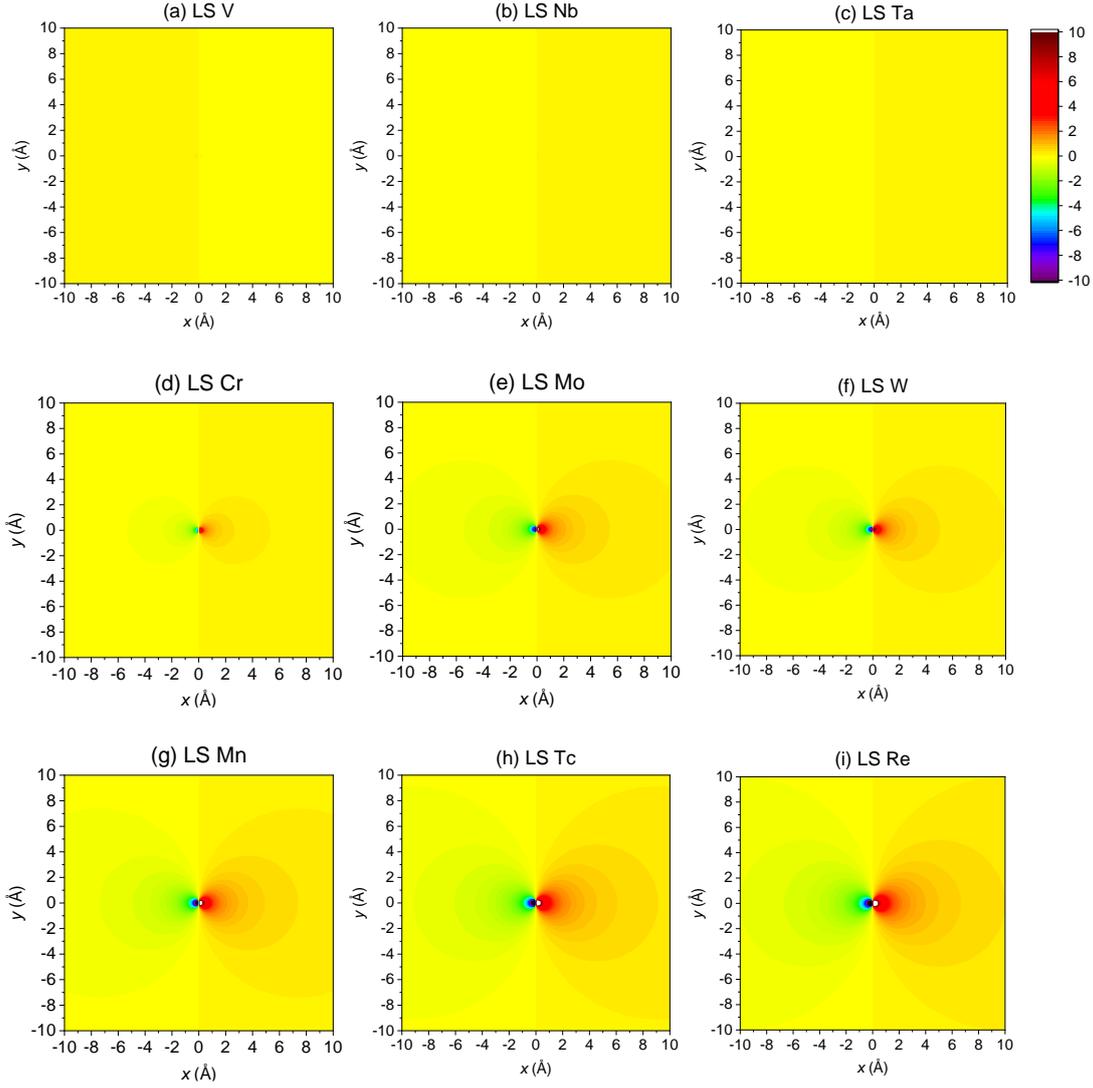

**Fig. 9.** Interaction energies (unit in eV) $E^{\text{int}}(x,y)$ between the solute atoms and prismatic plane ⟨a⟩ screw dislocations in α-Ti. The subfigures (a) to (i) are respectively for V, Nb, Ta, Cr, Mo, W, Mn, Tc, Re. The solute atoms are located at the LS off-center sites.

*3.5. Force between the solute atom and dislocation*

With the calculated interaction energy presented in Section 3.4, the forces $F_x$ between the SAs and the dislocations are evaluated by using Eq. 12. The calculated SA-dislocation interaction energy demonstrates that the SA may segregate above/below the slip plane (HS-SA above/below BPED and PPED as seen in Fig. 6 and Fig. 7 and LS-SA above/below BPSD as seen in Fig. 8) or on the slip plane ahead/behind the dislocation (LS-SA on the slip planes of PPED and PPSD as seen in Fig. 7 and Fig. 9).

Therefore, to evaluate the forces between the SAs and BPED, the SAs are placed both above the slip plane with a fixed height $h$ and on the slip plane with $h = 0$ (see Fig. 2) for comparison. We take $h = 2b$ to avoid the failure of the elasticity theory in the dislocation core region [46,47] for SA above the slip plane. Similarly, for SA on the slip plane, the forces between SA and dislocation are only valid with SA-dislocation distance out of the range from $-2b$ to $2b$.

3.5.1. Forces between SA and edge dislocations

The SAs tend to segregate above or below the slip plane of BPED (see Fig. 6). Therefore, the forces between the BPED and SAs are calculated with SAs above/below the slip plane with height $h = 2b$, which are displayed in Fig. 10. For the SAs on the slip plane of the BPED, i.e., $h = 0$, the forces are exactly zero. Therefore, we do not discuss it in this paper.

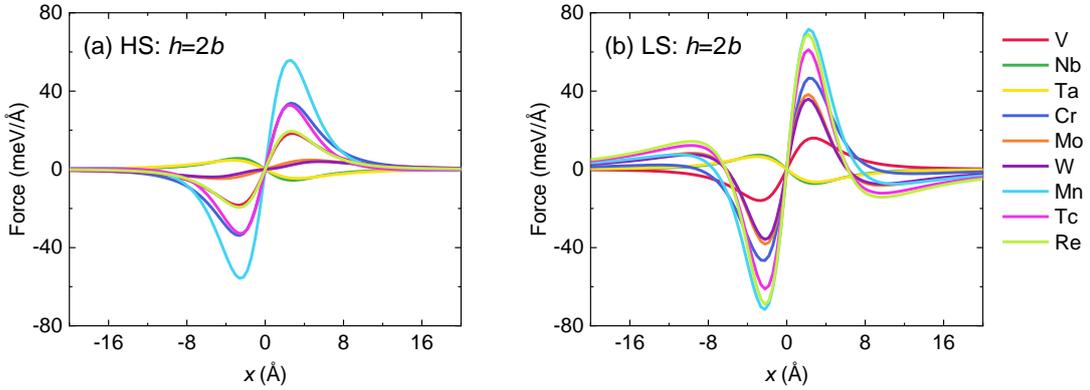

**Fig. 10.** Forces between the basal plane edge ⟨a⟩ dislocations and the solute atoms V, Nb, Ta, Cr, Mo, W, Mn, Tc, and Re against the location of the dislocation in its glide direction $x$. Subfigure (a) is for the solute atoms occupying the high symmetry lattice cite while (b) for the solute atoms at the low symmetry off-center site. The forces are evaluated with the solute atoms placed $h = 2b$ above the slip plane (see Fig. 2).

As seen in Fig. 10a, among all the SAs, Mn corresponds to the strongest maximum force $F_m$ between the HS-SA and BPED (about 56 meV/Å). $F_m$ decreases in the order of Mn, Cr/Tc, Re/V, Mo/W, Nb/Ta. The trend of $F_m$ against the HS-SAs is exactly in accordance with the lattice strain they induced as seen in Fig. 4a.

In general, $F_m$s for the LS-SAs are slightly stronger than those for the HS-SAs except for V, Nb, and Ta. The strongest $F_m$ is about 72 meV/Å for LS-Mn. $F_m$ decreases in the order of Mn, Re, Tc, Cr, Mo, W, V/Nb/Ta. Interestingly, such a sequence is slightly different from the trend of the main lattice strains $\lambda_1$ and $\lambda_2$ they induced. As seen in Fig. 4b, the main lattice strains $\lambda_1$ and $\lambda_2$ induced by the LS-SAs decrease in the order of Re, Tc, Mn, Mo, W, Cr, V/Nb/Ta as shown in Fig. 4b. The local lattice distortions along the $Z$ axis ($\lambda_3$ in Fig. 4b) might be responsible for the different trends of the interaction energy and the lattice strains $\lambda_1$ and $\lambda_2$. $\lambda_3$s caused by Cr and Mn are respectively larger than those by Mo/W and Tc/Re which are respectively in the same groups of Cr and Mn in CEPT, and, therefore, contribute more to the interaction energy.

For both the HS- and LS-SAs, $F_m$ occurs at $x$ (i.e., the distance between the SA and dislocation along the dislocation slip direction) about ±2 Å. The force between HS-SA and BPED vanishes with $x$ over the range of ±8 Å. However, for the LS-SAs, a secondary maximum force appears at $x \approx ±8$ Å and the force remains sizable with $x$ over the range of ±8 Å. Namely, the force between the LS-SA and BPED is longer ranged than that between HS-SA and BPED.

For the force between the SA and PPED, we consider two cases: one with SA located above the slip plane with height $h = 2b$ and the other with the SA located on the slip plane, i.e., $h = 0$ because the HS-SA forms atmosphere above/below the slip plane of PPED while the LS-SA forms atmosphere nearly on the slip plane of PPED. The forces are displayed in Fig. 11.

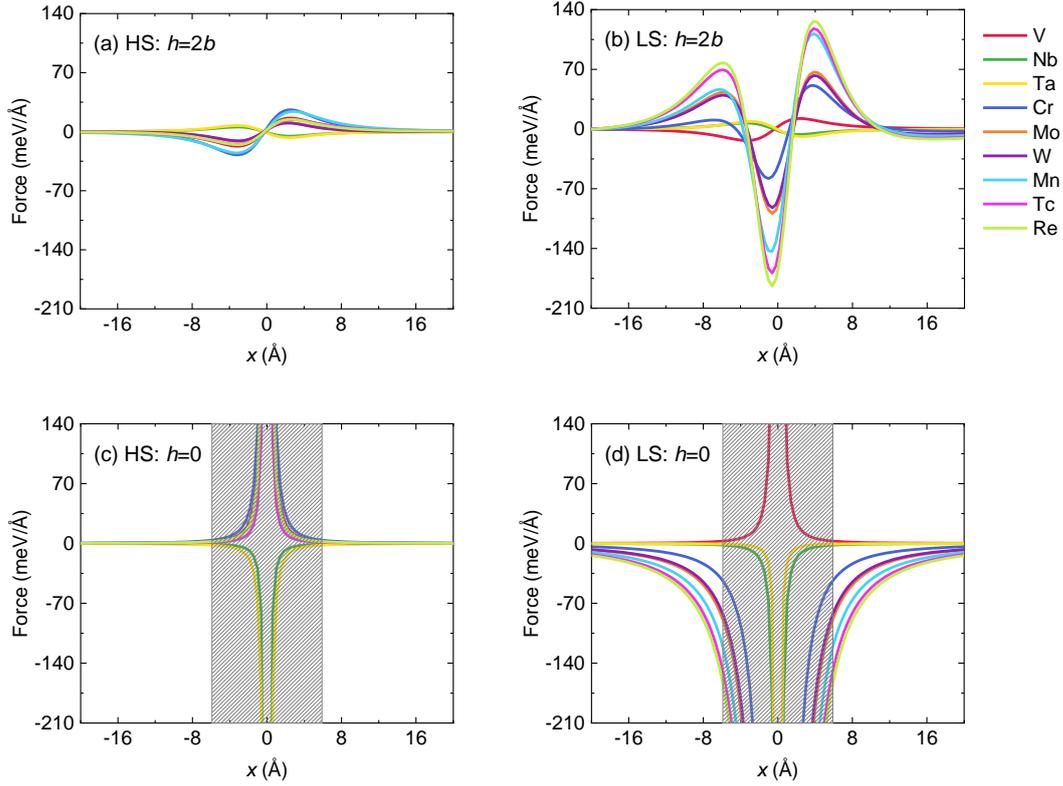

**Fig. 11.** Forces between the prismatic plane ⟨a⟩ edge dislocations and the solute atoms V, Nb, Ta, Cr, Mo, W, Mn, Tc, and Re against the location of the dislocation in its glide direction $x$. Subfigures (a) and (c) are for the solute atoms occupying the HS lattice cite while (b) and (d) for the solute atoms at the LS off-center site. In subfigures (a) and (b), the forces are evaluated with the solute atoms placed $h = 2b$ above the slip plane while in subfigures (c) and (d) they are evaluated with the solute atoms on the slip plane, i.e., $h = 0$ (see Fig. 2). In subfigures (c) and (d), the shadow from $x = -2b$ to $x = 2b$ represents the dislocation core.

If located above the slip plane with $h = 2b$, the HS-SAs exert relatively weak forces on the dislocation as seen in Fig. 11a. Among all the SAs involved in the present work, the strongest $F_m$ is about 27 meV/Å for Cr. $F_m$ occurs again at $x = \pm 2$ Å, similar to the case for the BPED. The force between the LS-SA is much stronger than that between the HS-SA except for V, Nb, and Ta. $F_m$ occurs at $x = 0$, which decreases in the sequence of Re, Tc, Mn, Mo, W, Cr, V/Nb/Ta, in agreement with the trend for the lattice strain induced by the LS-SAs. Secondary maximum forces appear at both negative and positive $x$ sides. For all the SAs, the force between LS-SA and PPED is not much longer ranged than that between HS-SA and PPED.

If located on the slip plane with $h = 0$, the HS-SAs impose essentially zero forces on the PPED outside the dislocation core area. However, the forces between the LS-SAs and the PPED are strong outside the dislocation core area, which decay smoothly with increasing distance between the SA and the dislocation. The sequence of $F_m$ against the SA is the same as that between the LS-SA and BPED. For both the LS- and HS-SAs, the huge force within the dislocation core area does not make sense because over there the elasticity theory for dislocation stress field becomes invalid.

3.5.2. Forces between SA and screw dislocations

Here, we focus ourselves on the forces between the LS-SA and the screw dislocations because the HS-SAs do not interact with screw dislocations.

The force between the LS-SA on the slip plane ($h = 0$) and BPSD is exactly zero, and, therefore, is not presented. Fig. 12 shows the forces between the LS-SAs and BPSD for $h = 2b$. The maximum force $F_m$ decreases with in sequence of Re, Tc, Mn, Mo, W, Cr, V/Nb/Ta. The strongest $F_m$ is about 75 meV/Å for Re. The forces are essentially zero for V, Nb, and Ta because the LS V/Nb/Ta is identical to the HS V/Nb/Ta.

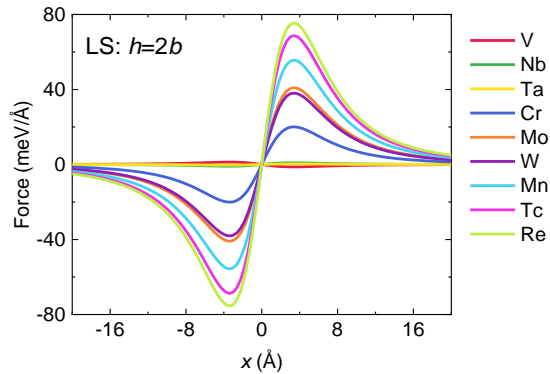

**Fig. 12.** Forces between the basal plane ⟨a⟩ screw dislocations and the solute atoms V, Nb, Ta, Cr, Mo, W, Mn, Tc, and Re against the location of the dislocation in its glide direction $x$. The solute atoms occupy the LS off-center site. The forces are evaluated

with the solute atoms placed $h = 2b$ above the slip plane (see Fig. 2).

The calculated interaction energy between LS-SA and PPSD suggests that the LS-SA tends to form atmosphere on the slip plane of PPSD. However, if located above/below the slip plane, the LS-SA may also impose force on the PPSD as the dislocation glides along the $x$ direction. Fig. 13a shows that $F_m$ between the LS-SA placed $h = 2b$ above the slip plane and the PPSD occurs at $x = 0$ and the force vanishes rapidly with increasing distance between the SA and dislocation. The strongest $F_m$ is about 116 meV/Å, again, for Re. Fig. 13b displays the forces between the LS-SAs on the slip plane and the PPSD. The forces decay with increasing distance between the SA and dislocation. The largest $F_m$ immediately out of the dislocation core area is about 116 meV/Å for Re, same as the corresponding maximum force for Re placed at $h = 2b$.

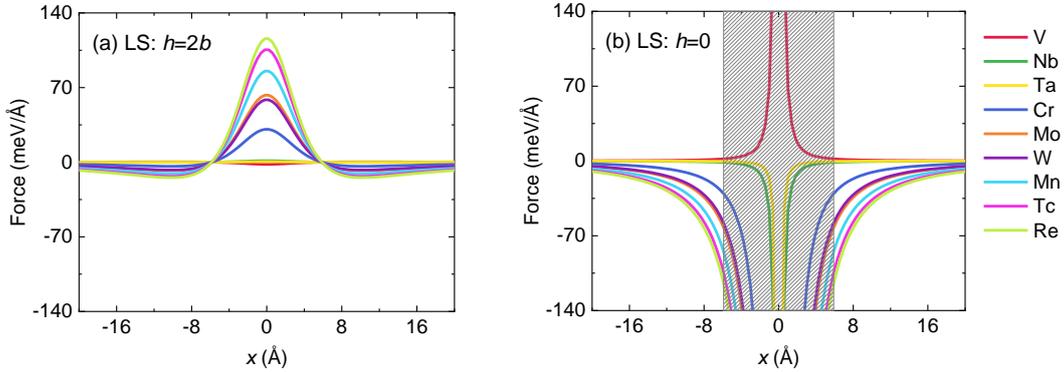

**Fig. 13.** Forces between the prismatic plane ⟨a⟩ screw dislocations and the solute atoms V, Nb, Ta, Cr, Mo, W, Mn, Tc, and Re against the location of the dislocation in its glide direction $x$. The solute atoms occupy the LS off-center site. In subfigures (a), the forces are evaluated with the solute atoms placed $h = 2b$ above the slip plane while in subfigures (b) they are evaluated with the solute atoms on the slip plane, i.e., $h = 0$ (see Fig. 2). In subfigures (b), the shadow from $x = -2b$ to $x = 2b$ represents the dislocation core.

*3.6. Solid solution hardening*

To quantify the contribution of the SA to the SSH of the α-Ti solid solution, the critical resolved shear stress (CRSS) increment $\Delta\tau$ induced by the SA is evaluated with the Labusch model (Eq. 13). For comparison, we adopt $F_m$s calculated with the SAs placed with height $h = 2b$ above the slip plane for all the SAs and various types of dislocations. For simplicity, the Schmid factor $S_F$ is taken as its maximum value 0.5. Considering the limited solid solubility of transition group elements in titanium, we take the concentration $c$ of SA as 0.4 at. %. The SSH induced by SA on the slip plane (i.e., $h = 0$) is somehow tricky which will be discussed in Section 4. Fig. 14 displays $\Delta\tau$ due to the interactions between the SAs and various types of dislocations. The CRSS increments induced by SAs located on HS and LS sites are denoted respectively as $\Delta\tau_{HS}$ and $\Delta\tau_{LS}$.

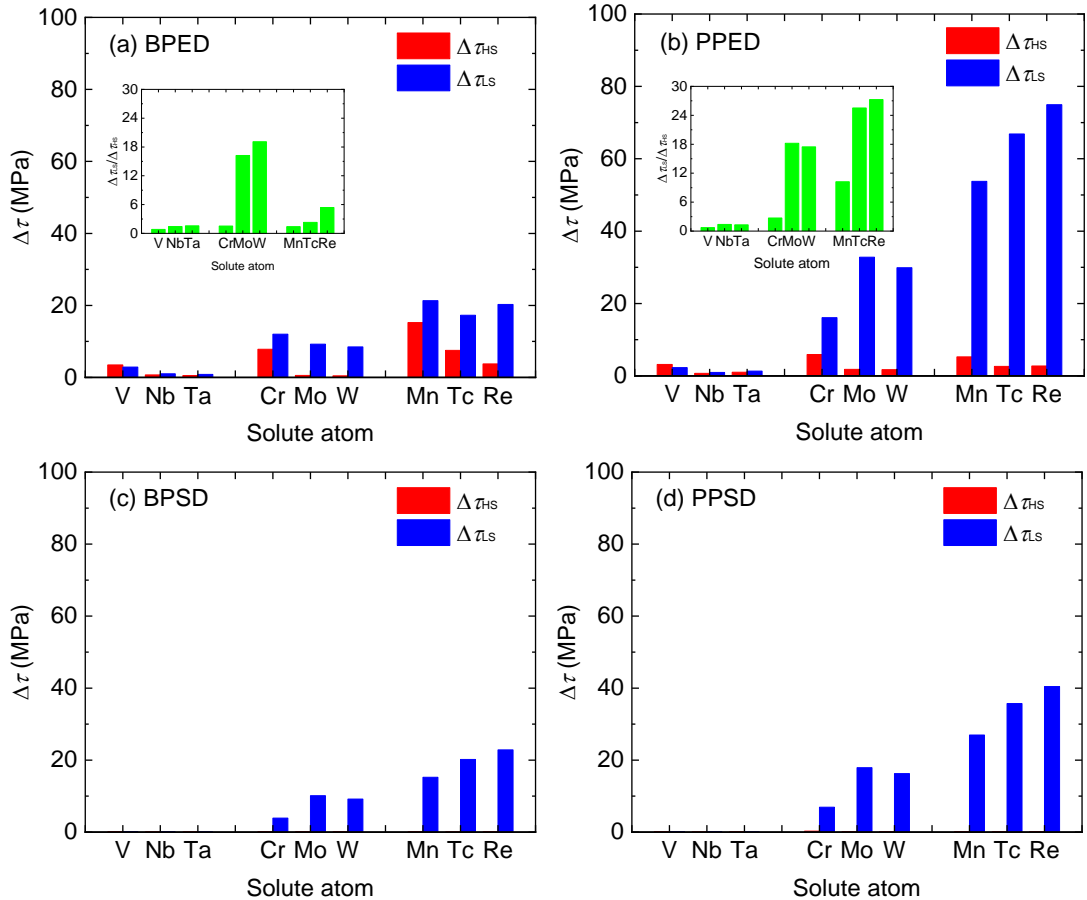

**Fig. 14.** Contribution of solute atom (SA = V, Nb, Ta, Cr, Mo, W, Mn, Tc, and Re) to the critical resolved shear stress, $\Delta\tau$ (MPa), evaluated with the interaction between SA and the basal plane edge dislocation (a), prismatic plane edge dislocation (b), basal

plane screw dislocation (c), and prismatic plane screw dislocation (d) with height $h = 2b$ above the slip plane. Red color is for $\Delta\tau$ induced by SA located at HS lattice site (denoted as $\Delta\tau_{HS}$) and blue for $\Delta\tau$ induced by SA at LS off-center site (denoted as $\Delta\tau_{LS}$). The inset in (a) and (b) shows the ratios $\Delta\tau_{LS}/\Delta\tau_{HS}$.

For BPED, $\Delta\tau_{HS}$ is visible for V, Cr, Mn, Tc, and Re (Fig. 14a) because these atoms result in sizable local lattice strain in α-Ti even located at HS site (Fig. 4a) due to their large atomic mismatch with the host Ti atom. The largest $\Delta\tau_{HS}$ is about 15 MPa for Mn. $\Delta\tau_{HS}$ for Nb, Ta, Mo, W is negligibly because they have atomic sizes close to that of the host atom Ti such that the local lattice strain is trivial. $\Delta\tau_{LS}$ induced by LS-V/Nb/Ta/Cr/Mn/Tc/Re is almost at the same level as the corresponding $\Delta\tau_{HS}$. The largest $\Delta\tau_{LS}$ is about 21 MPa for Mn. For most of the SAs, the $\Delta\tau_{LS}/\Delta\tau_{HS}$ ratio is small as shown in the inset of Fig. 14a except for Mo and W. Although $\Delta\tau_{LS}$ is not large for Mo and W, $\Delta\tau_{LS}/\Delta\tau_{HS}$ is still more than 16 due to the trivial $\Delta\tau_{HS}$. For both the HS and LS, the largest $\Delta\tau_{LS}$ is no more than 22 MPa.

For the PPED, the HS-SAs induced $\Delta\tau_{HS}$ is not significant for all the SAs, similar to the case for BPED. The LS-SAs induced $\Delta\tau_{LS}$ is much larger than that induced by the corresponding SAs on HS site except for V, Nb, and Ta. The strength increments $\Delta\tau_{LS}$ for Tc, and Re reach respectively to about 67 and 75 MPa, more than 25 times of the corresponding $\Delta\tau_{HS}$.

The HS-SAs do not contribute to $\Delta\tau$ through interacting with the screw dislocations (Fig. 14c and 14d), i.e., $\Delta\tau_{HS} \approx 0$. This is also true for LS-V/Nb/Ta as they are identical to HS-V/Nb/Ta. However, for LS-Cr/Mo/W/Mn/Tc/Re, $\Delta\tau_{LS}$ is considerably larger. Namely, the solid solution may be strengthened due to the drag of the screw dislocation by these solutes.

## 4. Discussion

Recently, our first principles calculations demonstrated that some substitutional SAs in α-Ti solid solution prefer to occupy the LS position away from the HS lattice site. This finding is profound and renew our textbook knowledge about substitutional metal solid solutions and is expected to bring some new understanding on the properties of the solid solution such as SSH. Motivated by the finding, we investigated the interaction between the SA on the LS off-center site and the dislocation in α-Ti, in comparison with the one between the SA on the HS lattice site and the dislocation. The SSH induced by the SA was also evaluated. Indeed, the work provides some new insights into the substitutional metal solid solutions and SSH theory.

*4.1. Formation of solute atmosphere around dislocation*

According to Cottrell [5,6,48,49], to relieve the hydrostatic stresses $(\sigma_{11} + \sigma_{22} + \sigma_{33})/3$ around the dislocation, oversized SA tend to gather in dilated regions while undersized in compressed regions above/below the slip plane of an edge dislocation. The SAs gathering above/below the slip plane is the well-known Cottrell atmosphere. Our calculation confirms the formation of Cottrell atmosphere of the HS-SAs around the edge dislocations. As shown in Fig. 6a~i and Fig. 7a~i, the strongest interaction between the HS-SA and the edge dislocation occurs at the region above/below the slip plane right over/under the dislocation such that the solute atmosphere forms over there. For the LS-SAs, the atmosphere still forms above/below the slip plane of the BPED but not right over/under the dislocation (see Fig. 6e′~i′). However, for the PPED, the atmosphere locates nearly on the slip plane ahead/behind the dislocation along its glide direction (see Fig. 7d′~i′). Obviously, the LS-SA atmosphere around the edge dislocation is distinct from the conventional Cottrell atmosphere.

Furthermore, the substitutional SA has not been supposed to form any atmosphere around the screw dislocation [50] because of the zero SA-dislocation interaction according to Cottrell's theory. Our work demonstrated that there exists interaction between the LS-SA and the BPSD and PPSD. For the BPSD, the solute atmosphere occurs above or below the slip plane right over or under the dislocation (see Fig. 8). For PPSD, the atmosphere appears on the slip plane ahead or behind the dislocation (see Fig. 9). The atmosphere of the LS-SA atoms is similar to the Snoek atmosphere formed with the interstitial atoms such as C, N, and O around the screw dislocation [51–53].

The unique segregation behavior of the LS-SAs around the dislocation, different from the normal HS ones as discussed above, provides the possibility to verify indirectly the existence of the LS-SAs by observing the atmosphere around the dislocation using experimental techniques such high resolution transmission electron microscopy (HRTEM) [54,55].

*4.2. Solid solution hardening and its effect on slip system selection*

The present work brings about some new insights into the SSH effect in the substitutional metal solid solutions. On the one hand, for the edge dislocation, we demonstrated that, in general, the LS-SAs contribute more to the SSH effect than the HS-SAs (see Fig. 14a and 14b). On the other hand, we showed that the LS-SA may interact with the screw dislocations in α-Ti solid solution, which provides additional SSH of the solid solution besides the SSH induced by the interaction between SA and edge dislocation. The above two factors suggest that the conventional Cottrell theory assuming the HS occupation of the solute atom underestimates significantly the SSH effect of the substitutional solute atoms preferring the LS off-center site. For Mo that is commonly used as alloying element in Ti alloys, $\Delta\tau_{LS}$ is about 33 MPa for PPED and about 18 MPa for PPSD, which is already a strikingly strong SSH effect considering that the concentration of Mo is only 0.4 at.%, noting that the conventional Cottrell theory predicts CRSS increment of only about 1.8 MPa.

For pure α-Ti, the primary deformation mode is the prismatic $\langle a \rangle$ dislocation slip while the basal plane $\langle a \rangle$ dislocation slip is secondary. The CRSS for the prismatic $\langle a \rangle$ slip $\tau^P$ is about 181 MPa, 28 MPa less than that for the basal $\langle a \rangle$ slip $\tau^B$ [56], i.e., $\tau^P - \tau^B = -28$ MPa. As seen in Fig. 14, for both the edge and screw dislocations, $\Delta\tau_{LS}$ for the prismatic $\langle a \rangle$ dislocation is larger than that for the basal $\langle a \rangle$ dislocation for the same SA. Namely, the prismatic $\langle a \rangle$ dislocation is more strongly pinned by the LS-SA than the basal plane $\langle a \rangle$ dislocation. This makes it possible that the priority of the two slip systems be altered. Taking the Ti-(0.4 at.%)Mo solid solution as an example, $\Delta\tau_{LS}^P - \Delta\tau_{LS}^B = 24$ MPa for the edge dislocation (see Fig. 14a and 14b) assuming that Mo segregates equally around the basal and prismatic dislocations, which cancels out partly the difference between $\tau^P$ and $\tau^B$. Namely, the solute Mo atom makes the mobility of the basal $\langle a \rangle$ edge dislocation closer to that of the prismatic edge dislocation. Comparing Fig. 6e′ and Fig. 7e′, we find that the interaction between Mo and PPED is stronger than that between Mo and BPED. Thus, the Mo atoms segregates more strongly to the PPED than to the BPED, suggesting that in practice PPED is pinned by more Mo atoms than BPED, which further increases $\Delta\tau_{LS}^P - \Delta\tau_{LS}^B$. Thus, $\Delta\tau_{LS}^P - \Delta\tau_{LS}^B$ even possibly overrides $\tau^P - \tau^B$ and reverse the relative mobility of the BPED and PPED at certain Mo concentration. This effect is also true for the other LS-SAs such as W, Mn, Tc, Re. The HS-SAs are not expected to alter the relative mobility of the BPED and PPED because the difference between $\Delta\tau_{HS}^P$ and $\Delta\tau_{HS}^B$ is insignificant for the HS-SAs as seen in Fig. 14a and 14b. Notably, aluminum (Al), as a simple metal element without $d$ electrons, is expected to occupy the HS lattice site. However, the alloying of Al in α-Ti may change the priority of the prismatic and basal plane slips [57]. Nevertheless, such a change might not be ascribed to the elastic interaction between the dislocation and Al as addressed in the present work but to the interaction between Al and the dislocation core as indicated by the work of Yu et al. [58].

It should be noted that, for PPED and PPSD, the LS-SA forms atmosphere on the slip plane. If the atmosphere is behind the dislocation (i.e., dislocation gliding away from the atmosphere), the dislocation drags the atmosphere during its motion. The resulted SSH effect may still be described within the elasticity theory. However, if the atmosphere is ahead of the dislocation (i.e., dislocation moving toward the atmosphere), the SAs may finally enter the dislocation core region. In this case, the elasticity theory will fail to describe the SSH effect.

*4.3. Mechanism of SSH*

According to Cottrell's theory, the SSH effect is determined by the lattice strain due to the size mismatch between the substitutional SA and the matrix metal atom, which may be represented by the volume change $\Delta V$ induced by the SA assuming that the SA occupying the HS lattice site. $\Delta V$ may be conveniently calculated by using first principles methods as the volume difference between the optimized supercells of the pure Ti and Ti-SA solid solution. Fig. 15 displays the relationship between the CRSS increments induced by the SAs against the absolute value of $\Delta V$ for the edge dislocations.

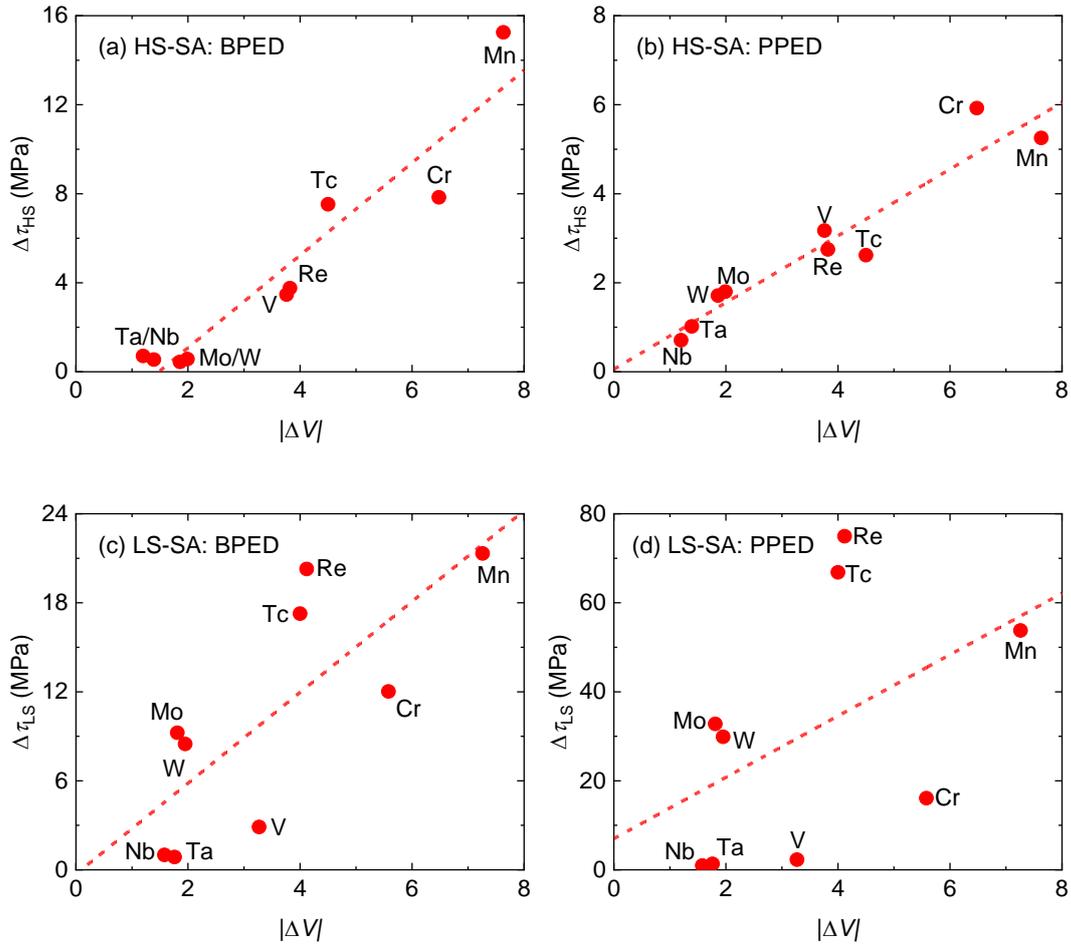

**Fig. 15.** Critical resolved shear stress increments $\Delta \tau$ against the absolute value of the volume change $|\Delta V|$ induced by the solute atom. Subfigures (a) and (b) are for $\Delta \tau_{HS}$s caused respectively by the interactions between basal and prismatic plane edge $\langle a \rangle$ dislocations (BPED and PPED) and the substitutional solute atom on HS lattice site while (c) and (d) for $\Delta \tau_{LS}$s caused respectively by the interactions between BPED and PPED and the substitutional solute atom at LS off-center site. The dash lines represent the linear fitting of the corresponding data points.

As seen in Fig. 15a and 15b, for both BPED and PPED, the HS-SA induced CRSS

$\Delta\tau_{HS}$ increases with $|\Delta V|$. $\Delta\tau_{HS}$ and $|\Delta V|$ exhibits fairly good linear relationship. The deviation of the data points from the exact linear relationship in Fig. 15a and 15b might be ascribed to the fact that, in the present work, $\Delta\tau_{HS}$ is calculated with the elastic dipole model where the lattice strain $\lambda_1 = \lambda_2 \neq \lambda_3$ due to the slightly anisotropic hcp lattice with the lattice parameter $c/a \approx 1.590$ deviating from the ideal value 1.633. $|\Delta V|$ does not reflect the anisotropy of the lattice.

Fig. 15c and 15d display the LS-SA induced CRSS increment $\Delta\tau_{LS}$ against $|\Delta V|$ for BPED and PPED. It is seen that $\Delta\tau_{LS} \sim |\Delta V|$ data points are very much scattered and do not show any definite relationship. As seen in Fig. 15c and 15d, LS-Re/Tc induces a relatively small $|\Delta V|$ but a significantly high $\Delta\tau_{LS}$ while LS-Cr causes a relatively large $|\Delta V|$ but a very low $\Delta\tau_{LS}$. The scattering of the $\Delta\tau_{LS} \sim |\Delta V|$ data points indicates that the volume change and the size mismatch between the substitutional SA and the host atom Ti may not account for the SSH effect for the LS-SAs, dissimilar to the case for the HS-SAs.

In the elastic dipole model, the shape of the dipole determines the SSH effect [38,59]. The shape of the dipole is in general described with the shape factor $|\lambda_1 - \lambda_2|$. In Fig. 16, we plot the LS-SA induced $\Delta\tau_{LS}$ against the shape factor $|\lambda_1 - \lambda_2|$. For all the dislocation types, $\Delta\tau_{LS} \sim |\lambda_1 - \lambda_2|$ exhibits excellent linear relationship, especially for PPED, BPSD, and PPSD. The shape factor $|\lambda_1 - \lambda_2|$ for the LS-SA is resulted from the split of $\lambda_1 = \lambda_2$ for the HS-SA when the SA shifts from the HS lattice site to the LS off-center site. As explained in our previous publications [26,27], the shift of the SA is ascribed to the Jahn-Teller splitting of the degenerated $d$ orbitals of the HS-SA in α-Ti. Therefore, $|\lambda_1 - \lambda_2|$ depends mainly on the strength of the Jahn-Teller splitting. A stronger Jahn-Teller splitting effect should results in a larger $|\lambda_1 - \lambda_2|$ and stronger SSH effect (i.e., larger $\Delta\tau_{LS}$). As seen in Table 1, the energy difference between the HS and LS configurations, $\Delta E$, for Mn, Tc, and Re (respectively 0.226, 0.311, and 0.295 eV) are larger than those for Cr, Mo, and W (respectively 0.065, 0.109, and 0.045 eV), indicating that Jahn-Teller splitting effect for Mn, Tc, and Re is stronger than that for Cr, Mo, and W. Therefore, $|\lambda_1 - \lambda_2|$ and $\Delta\tau_{LS}$ for Mn, Tc, and Re are larger than those for Cr, Mo, W.

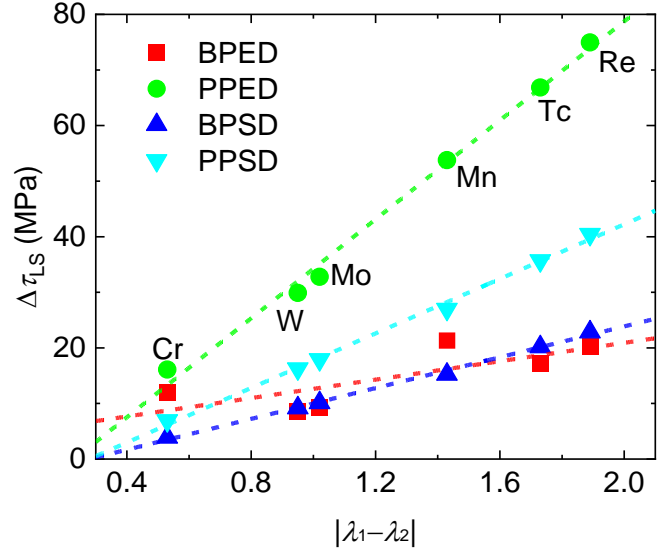

**Fig. 16.** Critical resolved shear stress increments $\Delta\tau_{LS}$ induced by the substitutional solute atom at LS off-center site against the shape factor of the elastic dipole $|\lambda_1 - \lambda_2|$. The red, green, blue, and cyan symbols are respectively for the basal and prismatic plane edge and screw ⟨a⟩ dislocations (BPED, PPED, BPSD, and PPSD). The dash lines represent the linear fitting of the corresponding data points.

## 5. Conclusion

Motivated by the finding about the off-center occupation of some substitutional solute atoms in α-Ti that is expected to enhance the SSH effect of the solid solution, we investigated the elastic interaction between the solute atom and dislocation in α-Ti using first principles calculations in combination with elastic dipole mode. The CRSS increment induced by the solute atom was subsequently evaluated with Labusch model. The main conclusions are listed as follows.

(1) In α-Ti, the substitutional solute atoms V, Nb, and Ta are stable to take the high-symmetry lattice site without changing the point group symmetry of the system whereas Cr, Mo, W, Mn, Tc, and Re shift to the low-symmetry off-center site.

(2) The substitutional solute atoms, if assumed to take the high-symmetry lattice sites, interact weakly with the basal and prismatic plane edge ⟨a⟩ dislocations in α-Ti and do not interact with the screw dislocations. The interactions between the edge dislocations and the solute atoms taking the low-symmetry off-center sites are in general stronger than those for the solute atoms on high-symmetry lattice site. Meanwhile, dissimilar to the solute atoms on the high-symmetry lattice sites, the solute atoms on the low-symmetry off-center sites interact strongly with the screw dislocation as well. The above two factors make the solid solution hardening effect induced by the low-symmetry off-center solute atom in general much stronger than that induced by the high-symmetry lattice site solute atom.

(3) The solid solution hardening effect induced by the low-symmetry off-center solute atom on the prismatic plane dislocation slip is stronger than the one on the basal plane dislocation slip, which makes it possible that the priority of the prismatic slip over the basal plane slip for pure α-Ti be reversed. The high-symmetry lattice site solute atom is not expected to change the priority of the slip systems within the framework of elastic continuum media theory.

(4) The solid solution hardening induced by the high-symmetry lattice site solute atom is mainly determined by the size mismatch between the solute and host atoms. The solid solution hardening induced by the low-symmetry off-center solute atom is dominated by the shape factor of the elastic dipole that is associated with the strength of Jahn-Teller splitting of the $d$ orbital of the solute atom.

(5) The interaction between the low-symmetry off-center solute atom and dislocation results in some unique segregation behavior of the solute atom around the dislocations. It forms atmosphere above/below the slip plane of the basal ⟨a⟩ dislocations but on the slip plane of the prismatic ⟨a⟩ dislocations regardless of the dislocation types (edge or screw), noting that conventional Cottrell theory predicts that substitutional solute atom only forms atmosphere above/below the slip plane of the edge dislocation but not form atmosphere around screw dislocation. The unique segregation behavior of the low-symmetry off-center solute atom provides the possibility to verify experimentally the low-symmetry off-center occupation of the solute atom through high resolution microscope observation.

**Acknowledgment**

The authors gratefully acknowledge the financial supports from the National Natural Science Foundation of China (grant Nos. U2106215, 52071315, and 52001307) and the National Key Research and Development Program of China (grant Nos. 2021YFC2801803 and 2022YFB3708300).